%% file: btou_nim.latex
\documentstyle{elsart}
\input{psfig}
\input{definitions.latex}

\begin{document}
\begin{frontmatter}

\title{A New Method for Selecting Exclusive Semileptonic Charmless 
$B$-Decays at $\eplemi$ Colliders at the $\Upsilon(4S)$}

\author[]{William S. Brower and Hans P. Paar\thanksref{hpp}}
\address{Physics Department 0319, University of California at San Diego,\break
9500 Gilman Drive, La Jolla CA 92093-0319}
\thanks[hpp]{Corresponding author, e-mail address: {\sf hpaar@ucsd.edu}}

\begin{abstract}
We introduce a new method for selecting exclusive semileptonic charmless 
$B$-decays in the presence of a large background.
The method can be applied to charged and neutral $B$-mesons decaying
into any exclusive neutral or charged hadronic final state.
The method is designed for high luminosity $\eplemi$ colliders operating 
at the $\Upsilon(4S)$.
It employs an {\sl improved} partial reconstruction technique for
$\Dstar$-mesons and a novel $0$-$C$ event fit to {\sl both} $B$-meson's decay 
products resulting in the kinematics of all particles (including neutrinos)
in the event.
The charged lepton energies are accessible from $1.0\, \GeV$ 
to the kinematic limit.
\end{abstract}
\end{frontmatter}

\section{Introduction}\label{sec:Intro}

Semileptonic $B$ decays into charmless final states provide
information about $|\Vub|$.
Although the relation between
$|\Vub|$ and the {\sl exclusive} charmless semileptonic branching ratio
is model dependent, this situation is improving as
lattice calculation results become available~\cite{ref:lattice}. 
In measuring the decay rates, it is desirable to sample as much of the
kinematically allowed phase space as possible.  
This increases signal efficiency and allows to differentiate between 
theoretical models. 

Branching fractions of exclusive semileptonic $B$ decays into 
non-charm final states have been measured\cite{ref:CLEObtopi}
to be approximately $2 \times 10^{-4}$.
Because this small rate leads to a relatively large background 
from $b\to c$ decays and other sources, 
we have developed\cite{ref:Brower} a new method with excellent
background rejection and acceptable signal efficiency. 
The method is based upon the kinematic
reconstruction of the entire event with one 
$B$-meson decaying into the low rate 
channel of interest (the ``signal'') and the other
$B$-meson into a channel with a large branching fraction (the ``tag'').
The method can only be used at $\eplemi$ colliders that 
operate at the $\Upsilon(4S)$ center-of-mass energy because the
kinematic reconstruction needs the magnitude of the $B$-meson's
momentum as an input.
If the beams have a non-zero crossing angle or unequal energy 
the method applies after 
a transformation of all observed final state particles
to the $\eplemi$ center of mass. 
The method can be used for charged and neutral $B$-mesons decaying into any 
exclusive neutral or charged hadronic final state.
The method requires large luminosities such as will become available at 
$B$-Factories.
The choice of a tag's decay channel 
is dictated by the need to have a large branching fraction
and a good detection efficiency.
The decay $B \ra \Dstar \ell \nuel$ 
has the largest branching fraction but 
it has the disadvantage of introducing a second neutrino in
addition to the neutrino on the signal side.  
As discussed in Sec.\,\ref{sec:EvtRec}, 
events containing two exclusive semileptonic $B$ decays can
be reconstructed if all particles' 3-momenta, except the two neutrinos, 
are measured.

To maintain efficiency, the $\Dstar$ is reconstructed using
an ``{\sl improved} partial reconstruction'' technique that,
like the ``{\sl standard} partial reconstruction'' technique,
only uses the pion kinematics from the decay $\Dstar \ra D \pi$,
see Sec.~\ref{sec:PartRec}.
The method is further improved by an novel
$0$-$C$ event fitting procedure, introduced in Sec.\,\ref{sec:EvtFit}.

\section{Event Reconstruction}\label{sec:EvtRec}

We consider the reconstruction of events of the 
type $B \ra X \ell^+ \nuel$ (signal), 
$\Bbar \ra Y \ell^- \nubarel$ (tag),
where the $B$, $\Bbar$, $X$, and $Y$ may be charged or neutral.
We assume that the momenta of the $X$, $\ell^+$, $Y$, and $\ell^-$ 
are measured.
Particle masses are assigned according to the signal hypothesis.
Incorrect assignments will generally cause the event to fail the procedure.
Thus the energies of the four measured particles are known.
The unknowns are the two angles associated with the direction
of the back-to-back $\BBbar$ and the momenta of the two neutrinos,
a total of 8.
Energy-momentum conservation for each $B$ decay gives 8 relations
so the kinematics of all particles in the event (including the two
neutrinos) can be calculated.
Solutions come in pairs due to square-roots in the equations.

\begin{figure}
\centering
\leavevmode
\psfig{figure=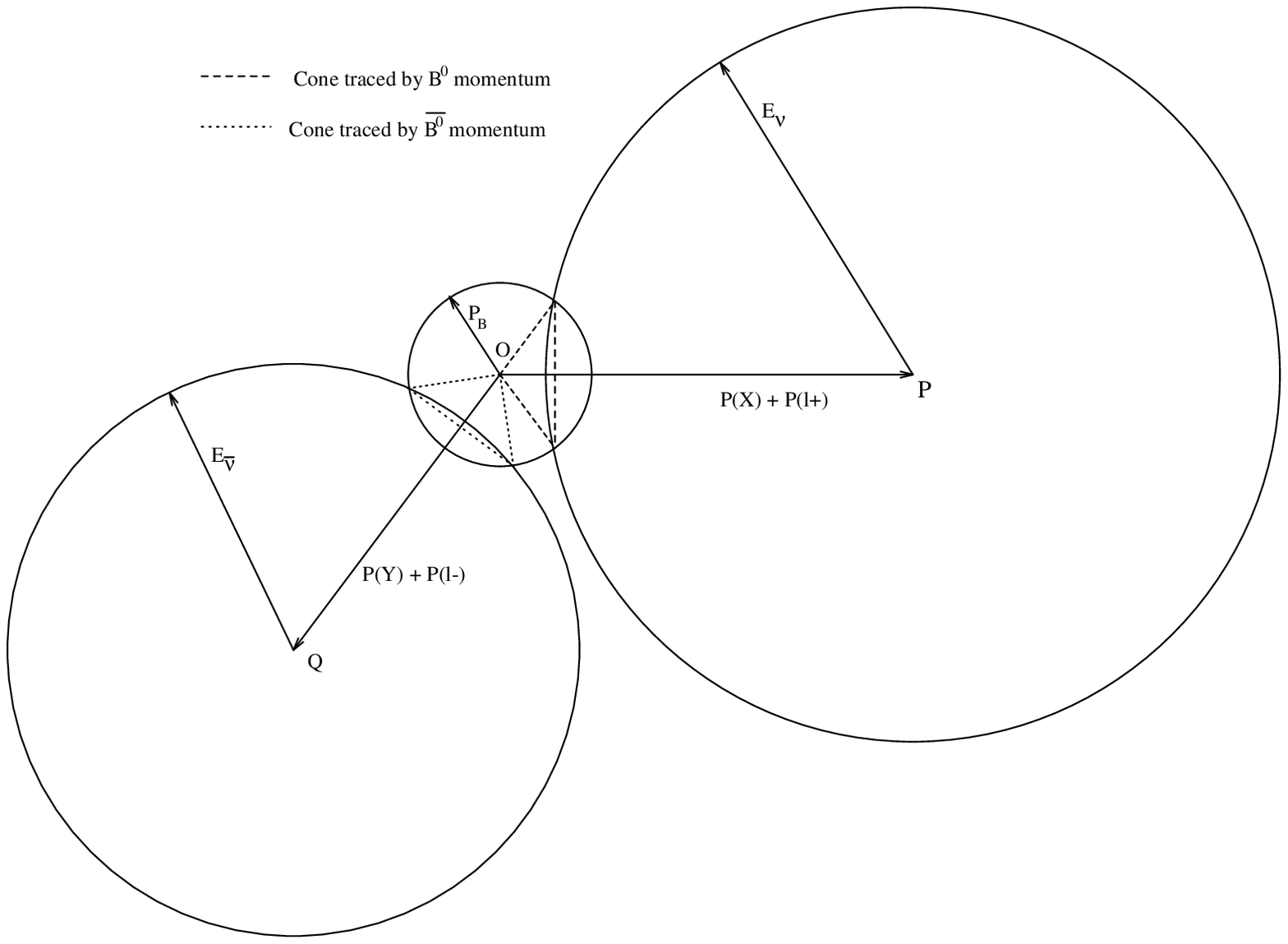,width=12.0cm}
\caption{Geometric construction that illustrates the event
reconstruction of $B \ra X \ell^+ \nuel$ and
$\Bbar \ra Y \ell^- \nubarel$.}
\label{fig:spheres}
\end{figure}

A geometric construction provides further understanding
of the method and demonstrates its powerful rejection 
of background events owing to three requirements, 
see Fig.\,\ref{fig:spheres}.
The endpoint of the momentum vectors of the back-to-back $B$-mesons
lie on the surface of a sphere of radius $325\, \MeV/c$
centered at their production point O (the ``$\BBbar$ sphere'').
Starting with the $B$ signal side, the sum of the momenta of 
$X$ and $\ell^+$ is calculated and shown as the vector OP.
Energy conservation applied to the $B$ decay gives the neutrino energy
$E_{\nuel} = E_{beam} - \sqrt{\vec{p}_{X}^{\,2} + m_{X}^2} - 
\sqrt{\vec{p}_{\ell^+}^{\,2} + m_{\ell}^2}$. 
The neutrino momentum vector must lie on a
sphere of radius $E_\nu$ centered on point P.
{\sl Constraint \# 1}: require that this sphere 
and the $\BBbar$-sphere intersect.
The intersection is a circle perpendicular to OP on which
the $B$ momentum vector is constrained to lie.
On the tag side, the sum of the momenta of $Y$ and $\ell^-$ 
is calculated and shown as the vector OQ.
The point Q is the center of a sphere with radius $E_{\nubarel}$, 
calculated from energy conservation applied to the $\Bbar$ decay.
{\sl Constraint \# 2}: require that this sphere and the $\BBbar$-sphere 
intersect.
The intersection is a circle perpendicular to OQ on which the endpoint
of the $\Bbar$ momentum vector is constrained to lie.
{\sl Constraint \# 3}: require that the $B$ and $\Bbar$ momentum vectors,
whose endpoints are constrained to lie on their respective circles,
be back-to-back.
To find the solution, reflect one of the circles through point O and find
the intersection of the reflected circle and the other circle.
If the circles intersect at all, there will be two intersections.
Background events will generally not satisfy all three constraints.

To maintain acceptable efficiency for signal events, 
the $\Dstar$ is not reconstructed from its $D \pi$ decay products: 
applicable branching fractions would cause an event loss of
an order of magnitude.
Instead the $\Dstar$ kinematics are inferred from the kinematics of the $\pi$
using the ``partial reconstruction'' technique.
To have the best possible determination of the $\Dstar$ kinematics
(the other three particles are measured with negligible errors)
for signal efficiency and background rejection we developed an 
{\sl improved} partial reconstruction technique, 
to be discussed next.
 
\section{Partial Reconstruction}\label{sec:PartRec}

Because 
the $\Dstar$-$D$ mass difference is barely large enough to create a pion, 
the decay products in the $\Dstar \ra D \pi$ decay have small momenta
in the $\Dstar$ center-of-mass system.
This feature is exploited in the long known partial reconstruction
technique\cite{ref:Fitch} to get a
{\em approximate} $\Dstar$ momentum using only the pion's kinematics.
Information on the $D$ is not needed.

\begin{figure}
\centering
\leavevmode
\psfig{figure=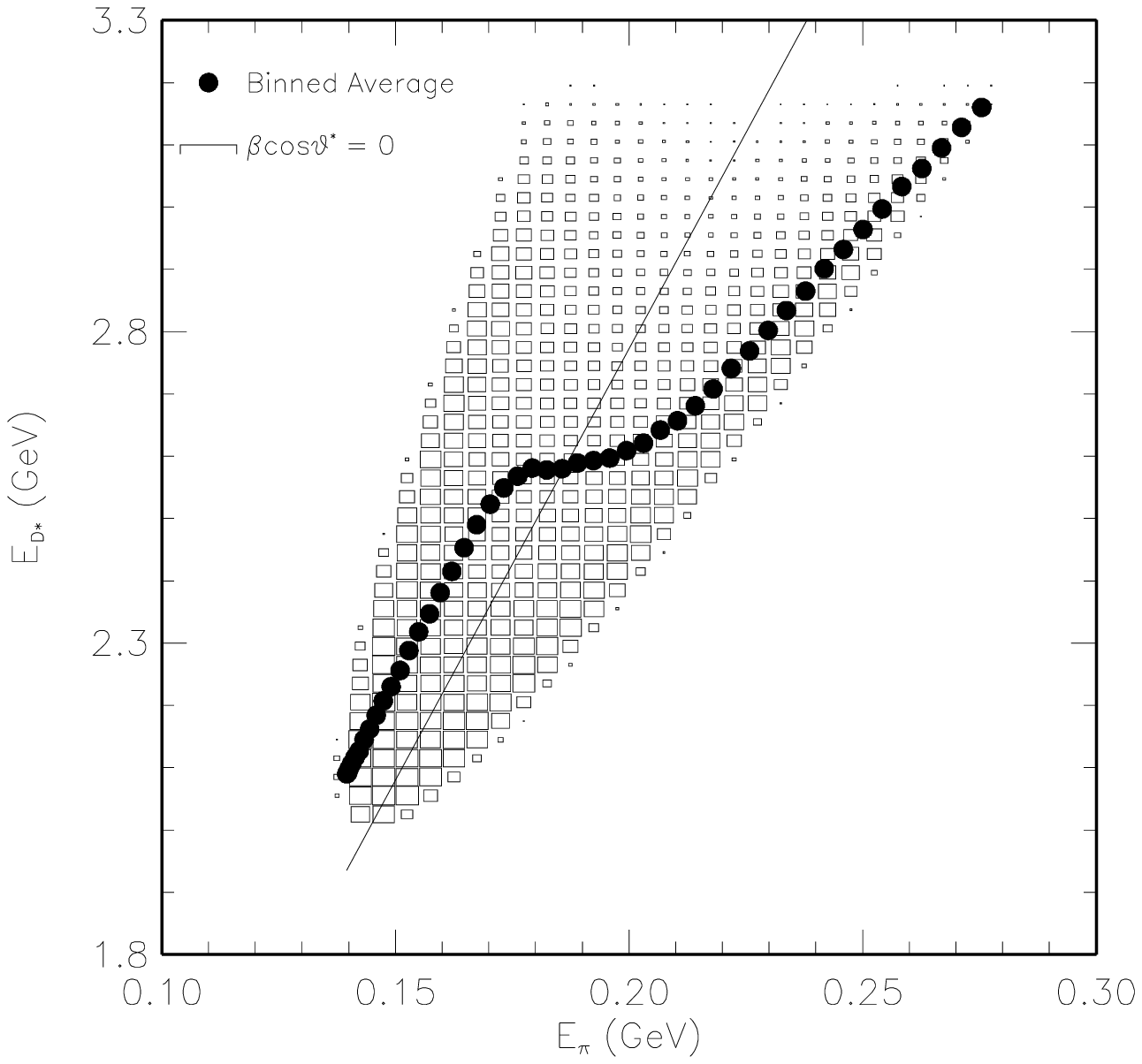,width=12.0cm}
\caption{Two-dimensional histogram of $E_{\Dstar}$ and $E_\pi$ 
in $\Bbarzero \ra \Dstarpl \ellmi \nubarel$, $\Dstarpl \ra \Dzero \pipl$. 
The dots are the average $\Dstar$ energy in each pion momentum bin. 
The line represents Eq.\, (\ref{eq:answerthree}).}
\label{fig:EDstar}
\end{figure}

The energy in the laboratory of the $\Dstar$ is given by
\begin{equation}
E_{\Dstar} = {m_{\Dstar} E_\pi \over E_\pi^*}
{1 \over 1 + \beta_\pi^* \beta_{\Dstar} \cos{\theta^*}}
\label{eq:answerone}
\end{equation}
and the angle $\theta$ between the $\Dstar$ and the $\pi$ by
\begin{equation}
\tan{\theta} = {E_\pi^* \sin{\theta^*} \over E_\pi}
{1 + \beta_\pi^* \beta_{\Dstar} \cos{\theta^*} \over
\cos{\theta^*} + \beta_{\Dstar} / \beta_\pi^*}
\label{eq:answertwo}
\end{equation}
Variables with a $\ast$ are to be evaluated in the $\Dstar$ center-of-mass
system.
In the {\sl standard} partial reconstruction technique one sets 
$\beta_\pi^* = 0$.
This gives 
\begin{equation}
E_{\Dstar} = {m_{\Dstar} E_\pi \over E_\pi^*}
\label{eq:answerthree}
\end{equation}
and $\theta = 0$.
A better approximation is obtained by studying  
the two-dimensional histogram of $E_{\Dstar}$ and $E_\pi$, 
see Fig.\,\ref{fig:EDstar},
for the case $\Bbarzero \ra \Dstarpl \ellmi \nubarel$, 
$\Dstarpl \ra \Dzero \pipl$. 
The ISGW\cite{ref:ISGW} model was used to simulate 
the decays.
The dots are the average $\Dstar$ energy in each bin of 
$p_\pi$.
The line represents Eq. (\ref{eq:answerthree}).
There is a significant difference between the dots and the line,
especially at larger energies.

\begin{figure}[htbp]
\begin{minipage}[t]{2.5in}
\psfig{figure=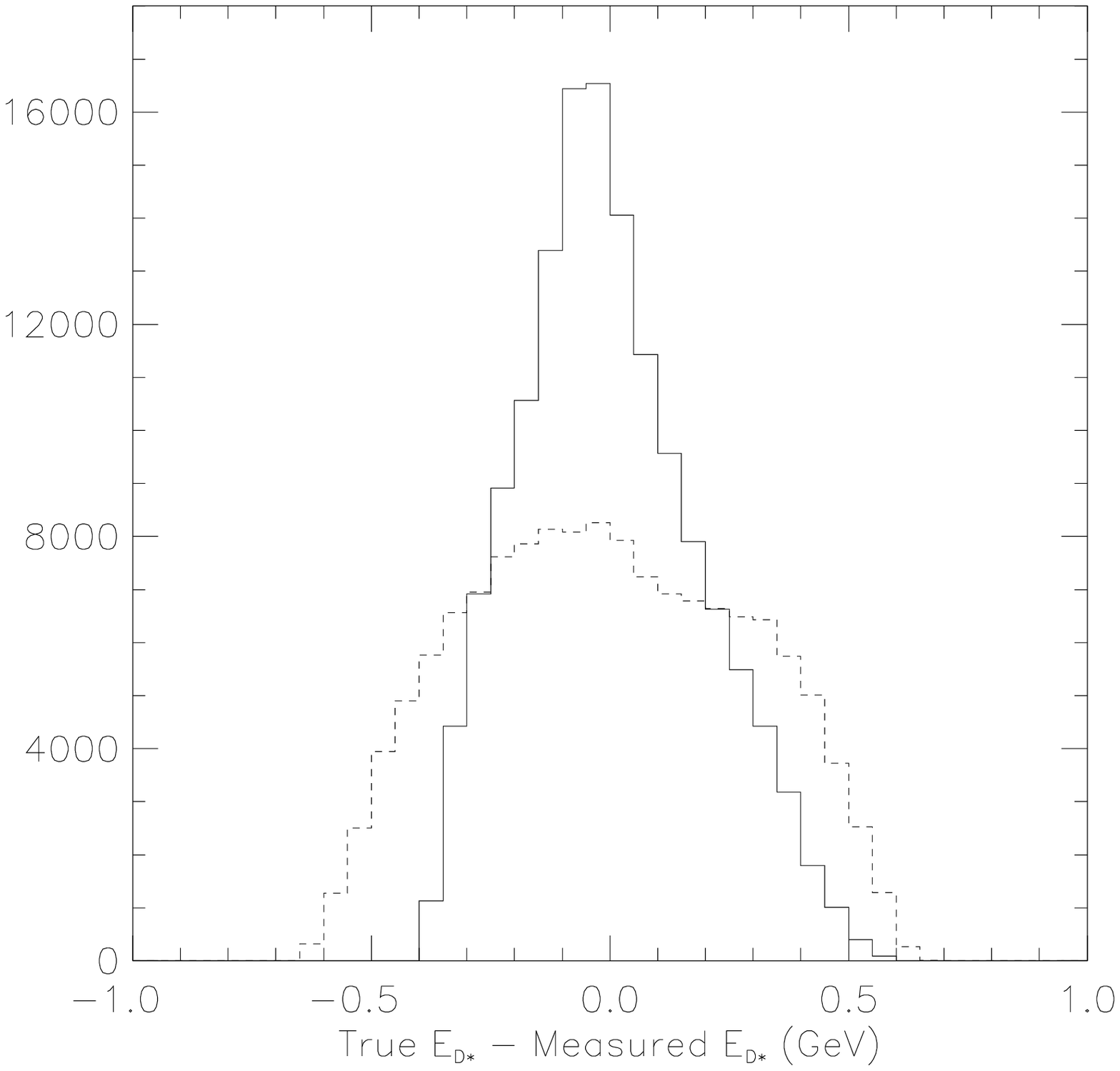,width=2.5in}
\caption{Difference between true $\Dstar$ energy and the 
$\Dstar$ energy from partial reconstruction. 
The dashed histogram is for the {\sl standard} and the solid histogram is for
the {\sl improved} partial reconstruction.}
\label{fig:deltaE}
\end{minipage}
\hfill
\begin{minipage}[t]{2.5in}
\psfig{figure=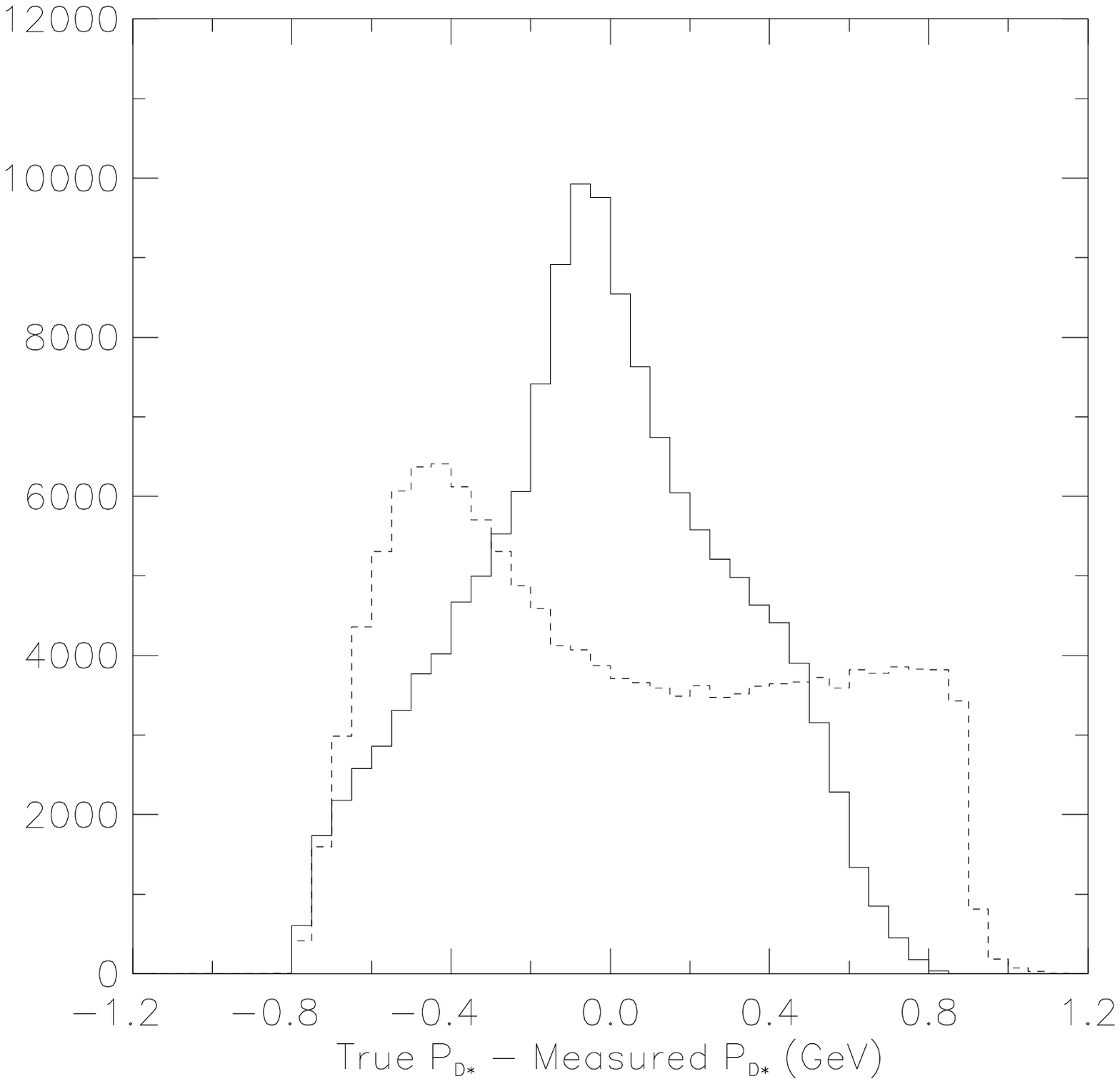,width=2.5in}
\caption{Difference between true $\Dstar$ momentum and the 
$\Dstar$ momentum from partial reconstruction. 
The dashed histogram is for the {\sl standard} and the solid histogram is for
the {\sl improved} partial reconstruction.}
\label{fig:deltap}
\end{minipage}
\end{figure}

A lookup table can be constructed to obtain $E_{\Dstar}$ from $E_\pi$ 
(or $p_\pi$).
We show in Fig.\,\ref{fig:deltaE} the difference between 
the true $E_{\Dstar}$ and $E_{\Dstar}$ obtained with the 
{\sl standard} partial reconstruction technique
(solid histogram) and the {\sl improved} partial reconstruction technique
(dashed histogram) that uses such a lookup table.
The improvement is clear.
The improvement can be seen even more clearly in the difference between 
the true $p_{\Dstar}$ and the partially reconstructed 
$p_{\Dstar}$, see Fig.\,\ref{fig:deltap}
The rms deviation of this distribution decreased from
$475\, \MeV/c$ to $330\, \MeV/c$ as a result of the improvement.
The distribution of $E_{\Dstar}$ energy in a given $E_\pi$ bin is 
very skewed and for some applications the most probable instead of the mean
may be more appropriate.
A similar method may be used to improve the estimated direction of the
$\Dstar$.
We found that the benefits from improving the $\Dstar$ direction 
are small.

In principle there is model dependence in 
the {\sl improved} partial reconstruction technique.
We have compared two models, ISGW\cite{ref:ISGW} and ISGW2\cite{ref:ISGWtwo} 
that both describe the $\Bzero \ra \Dstarpl \ell^- \nubarel$ and found the 
difference between them to be negligible relative to their respective 
difference from the {\sl standard} partial reconstruction.

\section{Event Fit}\label{sec:EvtFit}

\begin{figure}
\centering
\leavevmode
\psfig{figure=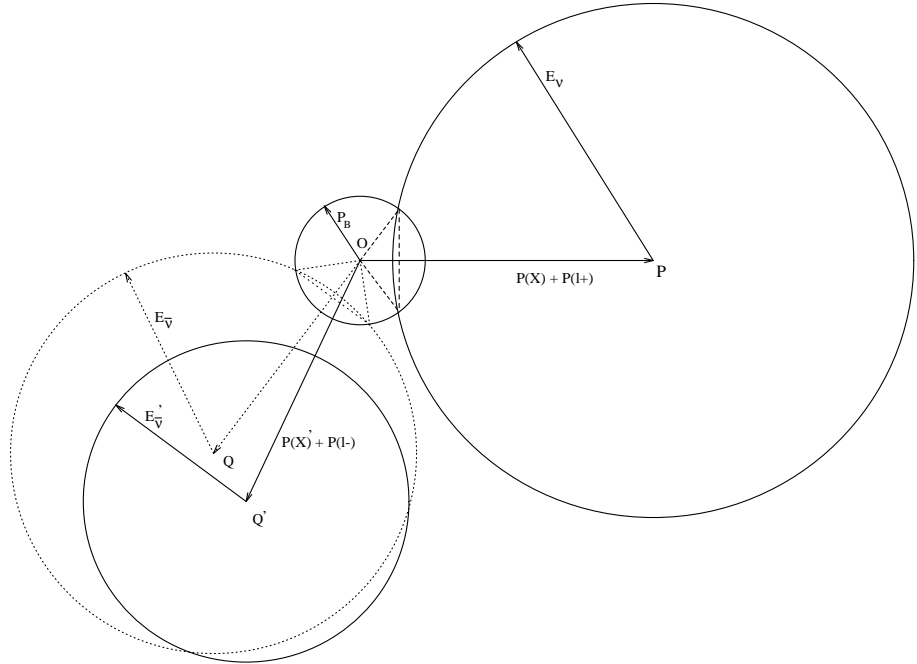,width=12.0cm}
\caption{The same event as in Fig.\,\ref{fig:spheres} showing that it 
now fails the event reconstruction because of the use of the
partial reconstruction technique for the $\Dstar$. 
Point Q (Q$\prime$) is the endpoint of the vector sum of the lepton and 
the true (partially reconstructed) $\Dstar$ momentum.}
\label{fig:badspheres}
\end{figure}

Because the $\Dstar$ momentum from {\sl improved} partial reconstruction 
has an uncertainty of approximately $330\, \MeV/c$, 
signal events may fail the event reconstruction.
An example is shown in Fig.\,\ref{fig:badspheres} where the
solid sphere centered on point Q$\prime$ is the result of the 
{\sl improved} partial reconstruction of the $\Dstar$
while the true $\Dstar$ leads to the dashed sphere centered on point Q.
This is the same event as shown in Fig.\,\ref{fig:spheres}.
Now the event does not satisfy constraint \#2 and would therefore
be rejected as being background.
It is possible to recover such signal events by varying the $\Dstar$ 
momentum (causing Q$\prime$ and $E_{\nubar}$ to vary as well) 
such that an intersection of the sphere centered on Q$\prime$
and the $\BBbar$ sphere exists (constraint \# 2) and that 
appropriate back-to-back $B$s exist (constraint \# 3), 
see Sec.\,\ref{sec:EvtRec}.
This corresponds to performing a $0$-$C$ fit.
Of all $\Dstar$ momenta that satisfy the three contraints,
we choose the one where the variation of the $\Dstar$ momentum
is minimal as measured by a $\chisq$, defined as follows.

Using a model to simulate $B \ra \Dstar \ell \nu\;,\; \Dstar \ra D^0 \pi$ 
we measured the covariance matrix $V_{ij}$ defined as
\begin{equation}
V_{ij} = {1 \over N} \sum_{\rm events} (p_{pr,i} - p_{tr,i}) 
(p_{pr,j} - p_{tr,j})\qquad i,j = x,y,z
\label{eq:covmat}
\end{equation}
where $p_{tr_i}$ ($p_{pr_i}$) is the $i$-th component ($i = x, y,
z$) of the true (partially reconstructed) $\Dstar$ momentum. 
Because there is significant dependence of the covariance matrix
upon the pion momentum, it is measured in bins of pion momentum. 
A $\chi^2$ is defined as
\begin{equation}
\chi^2 = \sum_i \sum_j (p_i - p_{pr,i}) (V^{-1})_{ij} 
(p_j - p_{pr,j})\qquad i,j = x,y,z
\end{equation}
where $p_i$ is the $i$-th component of the $\Dstar$ momentum.
The $\Dstar$ momentum with the lowest $\chi^2$ and
satisfying the three constraints is selected.
For signal events, $\chi^2_{min}$ is typically smaller than $1.0$ 
while background events generally have very large $\chi^2_{min}$.

\begin{figure}
\centering
\leavevmode
\psfig{figure=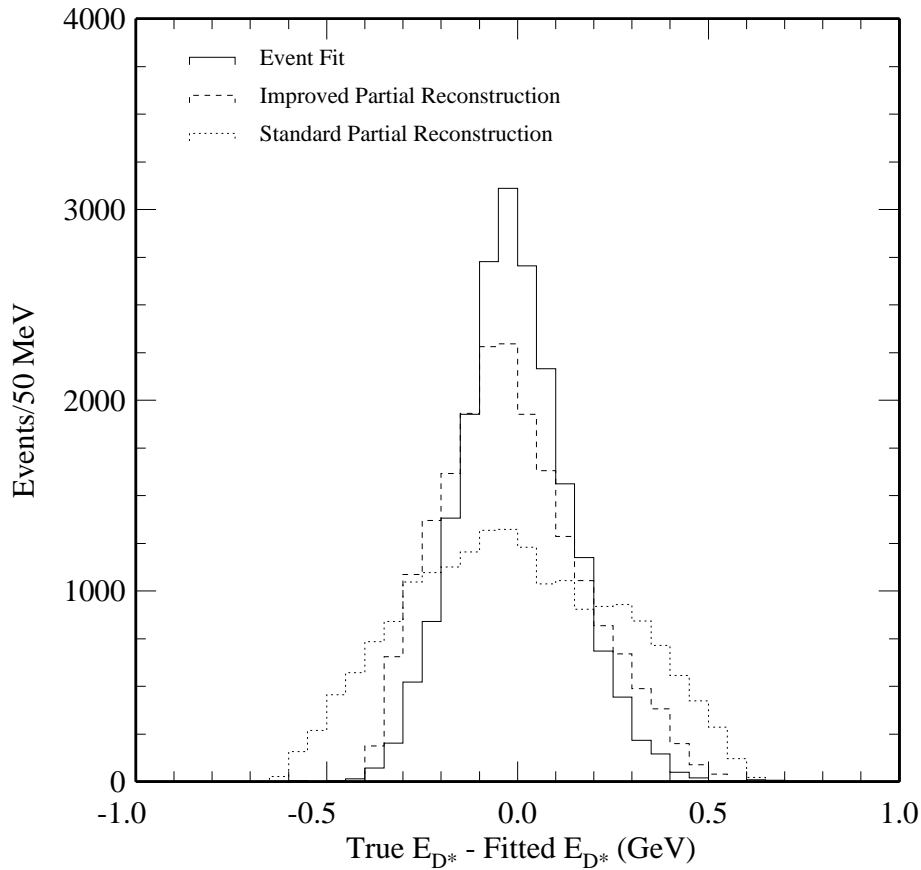,width=12.0cm}
\caption{Difference between the true and the fitted $\Dstar$ energy
(solid histogram).
Also shown are the difference between the true $\Dstar$ momentum and the 
$\Dstar$ momentum from respectively the {\sl improved} (dashed histogram) 
and the {\sl standard} (dotted histogram) partial reconstruction.} 
\label{fig:deltaE-search}
\end{figure}

In Fig.\,\ref{fig:deltaE-search} we show the difference between the 
true and the fitted $\Dstar$ momentum (solid histogram).
For comparison, we also show the difference of the true $\Dstar$ momentum 
and the $\Dstar$ momentum from respectively the {\sl improved}
(dashed histogram) and the {\sl standard} (dotted histogram) 
partial reconstruction.
Each histogram has the same number of events.
The fit significantly improves the $\Dstar$ kinematics, 
and that of its associated neutrino, 
thereby validating the fit procedure.

\section{Results}\label{sec:Results}

The event reconstruction technique has been evaluated
using simulated events.
The simulation uses the ISGW model to describe exclusive 
semi-leptonic $B$ decay and the detector simulation of the CLEOII 
detector\cite{ref:CLEODet}.
The simulation is known to describe the detector response well.
The simulation generates raw data in the same format as real data.
The simulated raw data are processed through the same analysis
procedures as the real data.
For definiteness we limit the discussion that follows to the reactions
$\Bzero \ra \pimi \ellpl \nuel$ (signal), 
$\Bbarzero \ra \Dstarpl \ellmi \nubarel$ (tag) with 
$\Dstarpl \ra \Dzero \pipl$.

A detailed discussion of the analysis is outside the scope of this paper.
Briefly, events with at least 4 charged tracks are selected, 
two of which must be leptons with momenta greater then $1.0\, \GeV/c$.
The leptons can have opposite or equal charges 
because of $\Bzero$-$\Bbarzero$ mixing.
At least one of the other two tracks must have a momentum between $40$ and
$200\, \MeV/c$, the kinematically allowed range for the pion from $\Dstarpl$
decay.
The most important background process is the one in which the signal $B$
also decays into a charmed final state.
To suppress this background at least partly, 
we calculate the effective mass and the total momentum 
of unused tracks and energy deposits 
in the electromagnetic calorimeter.
For signal events all unused tracks and energy deposits are the result of
$\Dzero$ decay so if their effective mass is greater than the 
$\Dzero$ mass the event is rejected.
Likewise, if the angle in the laboratory between their total momentum 
and the pion momentum is greater than $37\,{\rm deg}$, 
the event is rejected.

\begin{figure}[htb]
\centering
\leavevmode
\psfig{figure=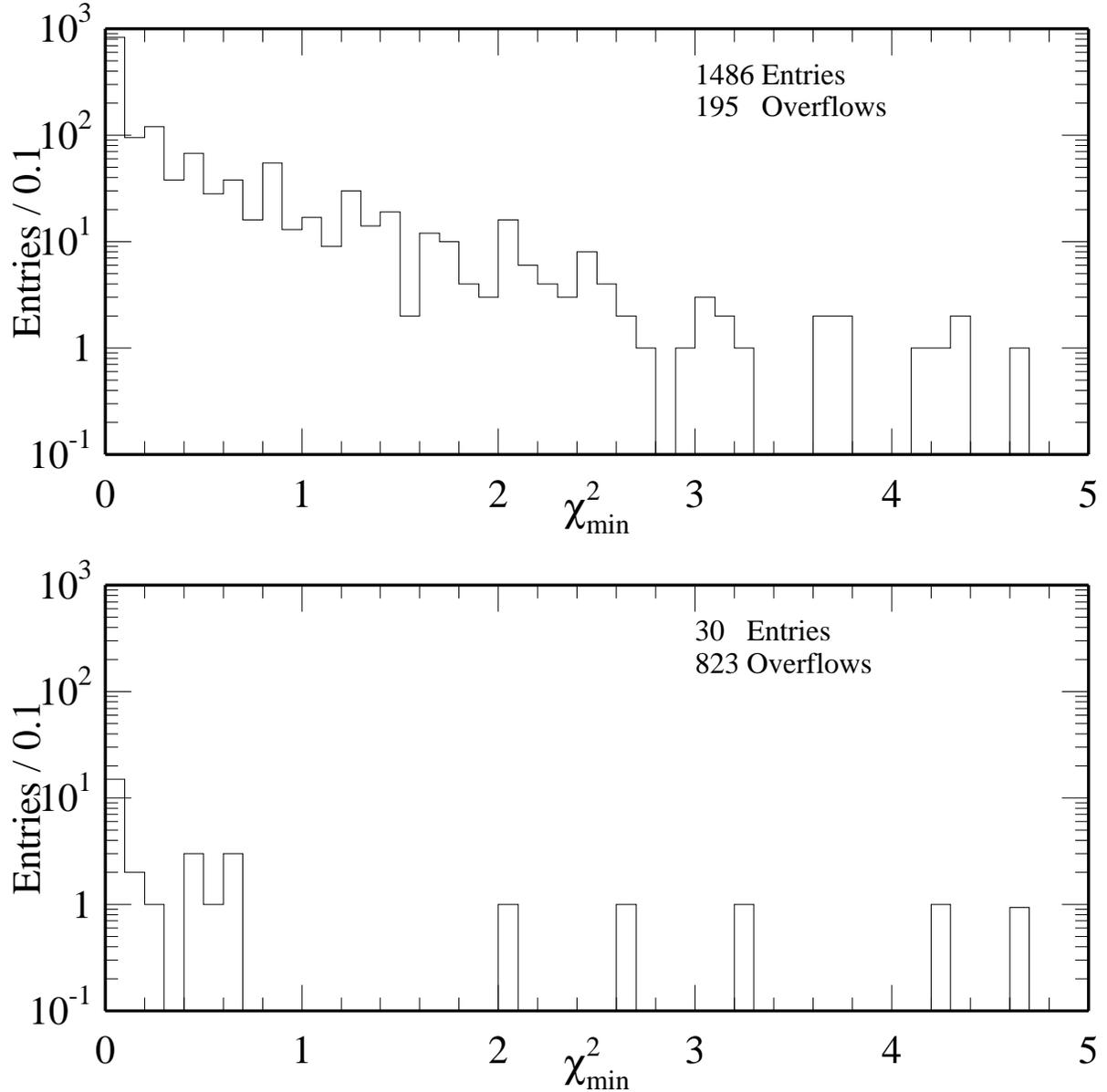}
\caption{$\chisqmin$ distributions for simulated signal (top) and 
background (bottom) events.
Note the logarithmic scale and the number of overflow events in each case.}
\label{fig:chisqmin}
\end{figure}

With this event selection, we loop over all combinations 
of four tracks in an event and over all events and apply the $0$-$C$
event fit and the {\sl improved} partial reconstruction.
The resulting $\chisqmin$ distribution is shown for signal 
and background events in Fig.\,\ref{fig:chisqmin}.
The latter are $\BBbar$ events that decay according to known decay modes, 
not including semi-leptonic charmless decays.
When we require $\chisqmin \le 1.0$, 
only 19 out of 853 background events pass, 
a rejection factor of 45.

The signal efficiency $\epsilon$, defined to include the branching fraction
$B(\Dstarpl \ra \Dzero \pipl ) = 0.68$, is measured to be 3.66\,\%.
The number of signal events is expected to be
\begin{equation}
N = 21 {\left(L \over 1\, \invfb\right)} 
\left( {B(\Bzero \ra \pimi \ellpl \nuel) \over 1.0 \times 10^{-4}}\right)
\, \epsilon 
\label{eq:Rate}
\end{equation}
where $L$ is the integrated luminosity.
The numerical factor includes the $\Bbarzero \ra \Dstarpl \ellmi \nubarel$
branching fraction and the $\BBbar$ production cross section.
$B$-factories aim for integrated luminosities of at least 
$30\, \invfb$/year.
A branching fraction 
$B(\Bzero \ra \pimi \ellpl \nuel) = 2 \times 10^{-4}$\cite{ref:CLEObtopi}
would result in about 50 events per signal channel with about 20 events
background.
When several signal channels are combined these numbers are expected 
to give measurements of $|\Vub|$ whose precision 
is dominated by theoretical uncertainties.
A vertex detector would improve the tracking of low momentum particles
(important for partial reconstruction) and allow the rejection of
some of the remaining charm background.
As indicated earlier, the method can be applied to charged and neutral 
$B$s decaying into any exclusive neutral or charged hadronic final states,
allowing many measurements of $|\Vub|$. 

{\bf Acknowledgements}

The authors thank their CLEO colleagues for the use of the CLEO analysis 
and event simulation code and the CLEO Monte Carlo dataset.
We have greatly benefitted from many fruitful discussions with colleagues
interested in semileptonic $B$-meson decays.

% Bibliography:

%\section*{References}

\end{document}

%% file: definitions.latex
\def\Journal#1#2#3#4{{#1} {\bf #2}, #3 (#4)}

\def\NIMA{{\em Nucl. Instr. Meth.} A}

\def\ra{\rightarrow}
\def\chisq{\chi^2}
\def\chisqmin{\chi_{\rm min}^2}

\def\ellpl{\ell^+}
\def\ellmi{\ell^-}
\def\nubar{\overline{\nu}}
\def\nuel{\nu_{\ell}}
\def\nubarel{\overline{\nu}_{\ell}}
\def\epl{e^+}
\def\emi{e^-}
\def\eplemi{\epl\emi}

\def\pipl{\pi^+}
\def\pimi{\pi^-}

\def\K0{K^0}

\def\invfb{\rm fb^{-1}}

\def\MeV{\rm MeV}
\def\GeV{\rm GeV}

\def\B0{B^0}
\def\B0s{B^0_s}%
\def\B0d{B^0_d}%

\def\Vub{V_{ub}}
\def\Dstar{D^{*}}
\def\Dstarpl{D^{*+}}
\def\Dzero{D^0}
\def\Bbar{\overline{B}}
\def\BBbar{B \overline{B}}                                                   
\def\Bzero{B^0}
\def\Bbarzero{\overline{B}^0}

%% file: btou_nim.bbl
\begin{thebibliography}{99}

\bibitem{ref:lattice}C. R. Allton {\it et al},
\Journal{Phys. Lett. B}{345}{513}{1995}; L. Lellouch, hep-ph/9609501 and
references therein.

\bibitem{ref:CLEObtopi}J.P. Alexander {\it et al}, (CLEO Collaboration),
\Journal{Phys. Rev. Lett.}{77}{5000}{1996}.

\bibitem{ref:Brower}W.S. Brower, PhD Thesis (1996), UC San Diego, 
La Jolla CA 92093-0319, unpublished.

\bibitem{ref:Fitch}Reference to first delta-m paper.

\bibitem{ref:ISGW}N. Isgur, D. Scora, B. Grinstein, and M.B. Wise (ISGW),
\Journal{Phys. Rev. D}{39}{799}{1989}.

\bibitem{ref:ISGWtwo}D. Scora and N. Isgur (ISGW2),
CEBAF Theory Group Preprint CEBAF-TH-94-14 (1994), hep-ph/9503486.

\bibitem{ref:CLEODet}Y. Kubota {\it et al} (CLEO Collaboration),
\Journal{\NIMA}{320}{66}{1992}.

\end{thebibliography}
